\documentclass[aps,prl,twocolumn,showpacs,superscriptaddress,groupedaddress]{revtex4}  
\usepackage{graphicx,comment,float,braket,textcomp,amsmath,lipsum,tikz}  
\usepackage{dcolumn}   
\usepackage{bm}        
\usepackage{amssymb}   
\usepackage{upgreek}

\usepackage{bbm}

\definecolor{mauve}{rgb}{0.5,0.5,0.9}
\definecolor{bblue}{rgb}{0,0.8,1}

\begin{document}

\title{Spectroscopic Signatures of Electron-Phonon Coupling in Silicon-Vacancy Centers in Diamond}

\author{Albert Liu}
\author{Steven T. Cundiff}

\affiliation{Department of Physics, University of Michigan, Ann Arbor, Michigan, 48109 USA}

\vskip 0.25cm

\date{\today}

\begin{abstract}
    Vacancy centers in diamond have proven to be a viable solid-state platform for quantum coherent opto-electronic applications. Among the variety of vacancy centers, silicon-vacancy (SiV) centers have recently attracted much attention as an inversion-symmetric system that is less susceptible to electron-phonon interactions. Nevertheless, phonon-mediated processes still degrade the coherent properties of SiV centers, however characterizing their electron-phonon coupling is extremely challenging due to their weak spectroscopic signatures and remains an open experimental problem. In this paper we theoretically investigate signatures of electron-phonon coupling in simulated linear and nonlinear spectra of SiV centers. We demonstrate how even extremely weak electron-phonon interactions, such as in SiV centers, may be completely characterized via nonlinear spectroscopic techniques and even resolved between different fine-structure transitions.
\end{abstract}

\pacs{}
\maketitle

\section{Introduction}

Electronic coupling to vibrations (phonons) is universal in solid-state systems. In solids of finite dimension, these interactions determine numerous properties such as bandgap renormalization \cite{Karsai2018,Liu2019-1} and electronic transport \cite{Giustino2017}. In zero-dimensional systems, such as vacancy centers in diamond, they play an equally important role. In nitrogen-vacancy (NV) centers, the presence of a large phonon sideband in their optical response comprises the primary limiting factor of zero-phonon line emission brightness \cite{Barclay2011}. Even in negatively-charged silicon-vacancy (SiV$^-$) centers, which are largely insulated from electron-phonon interactions by their inversion symmetric configuration, phonon-mediated dephasing processes still remain as the main limiting factor in their performance as quantum computing architectures \cite{Rogers2014-1,Sukachev2017}.

Due to its primary importance, characterizing the phonon spectral density $J(\omega)$ of such zero-dimensional systems has attracted sustained research interest. In systems comprised of discrete energy levels, $J(\omega)$ reflects the vibrational frequency-dependence of electron-phonon coupling and determines lineshapes in optical spectra \cite{Mukamel1999,Nitzan2006}. Traditionally the spectral density has been experimentally measured by mainly fluorescence line-narrowing (FLN) or spectral hole-burning (SHB), which are linear and third-order nonlinear spectroscopic techniques respectively. The main advantage of FLN and SHB is their ability to extract a homogeneous response from an inhomogeneously-broadened resonance via narrow-band excitation. Recently, multi-dimensional coherent spectroscopy (MDCS) was demonstrated to be a powerful method for directly measuring the ensemble-averaged spectral density with outstanding signal-to-noise ratio \cite{Liu2019-2}. In contrast to SHB, which measures the real-quadrature third-order optical response at a single excitation frequency, MDCS is capable of measuring the entire complex third-order optical response function $S^{(3)}(\omega_\tau,\omega_T,\omega_t)$.

However, characterization of the spectral density $J(\omega)$ is most commonly performed on optical spectra dominated by vibrational lineshapes such as molecular \cite{Pieper2011,Kell2013} or nanocrystal \cite{Palinginis2003,Liu2019-2} systems. In this paper, we perform a detailed analysis of how weak electron-phonon coupling manifests in different spectroscopic measurements (namely linear fluorescence and nonlinear transient-absorption and MDCS spectroscopies) of SiV$^-$ centers. We discuss how mixed time- and frequency-domain spectroscopies enable the complete characterization of the phonon spectral density in SiV$^-$ centers by relating electron-phonon coupling parameters to spectroscopic observables. In particular, we theoretically demonstrate how MDCS enables the measurement of vibronic coherence times in SiV$^-$ centers with exceptional sensitivity. We then emphasize the information uniquely obtained via nonlinear MDCS techniques.

\section{Theoretical Methods}

\begin{figure}[t]
    \centering
    \includegraphics[width=0.5\textwidth]{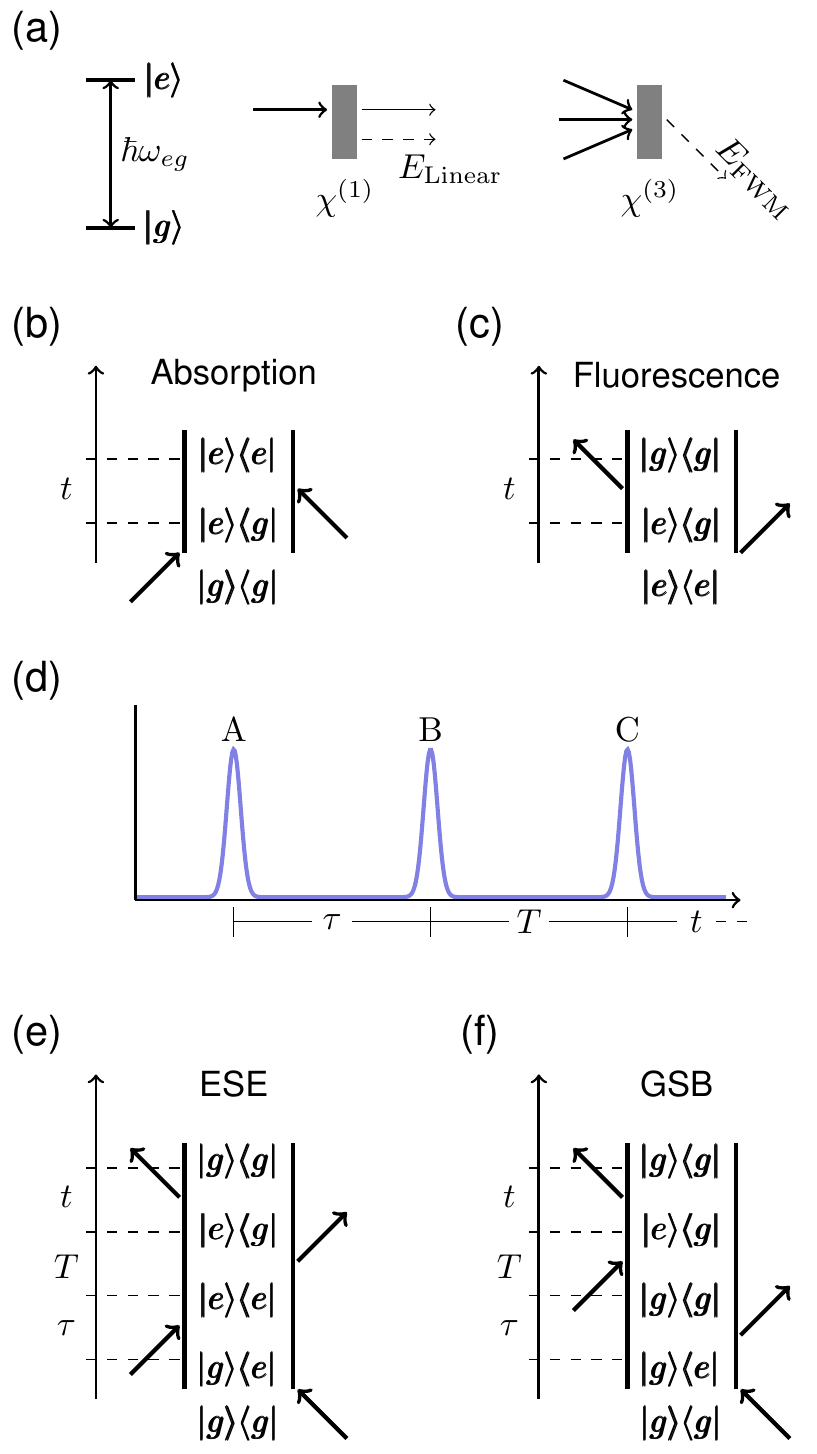}
    \caption{(a) Diagrams of (left) a two-level system and (right) linear/FWM spectroscopies. (b,c) Feynman diagrams of linear (b) absorption and (c) fluorescence. (d) FWM pulse-ordering diagram. Discussion in main text analyzes the FWM signal collected in the $-{\bf k}_{\text{A}} + {\bf k}_{\text{B}} + {\bf k}_{\text{C}}$ direction, in which this pulse-ordering generates a rephasing signal. (e,f) Feynman diagrams of (e) excited-state emission and (f) ground-state bleach third-order quantum pathways. We note that excited-state absorption pathways \cite{Hamm2011} are not possible for the two-level system considered here.}
    \label{Fig1}
\end{figure}

We begin with the simplest case of an optical transition between two energy levels $\Ket{g}$ and $\Ket{e}$ defined by an energy splitting $\hbar\omega_{eg}$ (shown in Fig.~\ref{Fig1}a). The interaction of this transition with a resonant optical field is defined by its optical susceptibility $\chi$ \cite{Boyd2008}, and for weak excitation intensities the optical susceptibility may be described perturbatively. For centrosymmetric systems the lowest-order terms are the linear and third-order optical susceptibilities $\chi^{(1)}$ and $\chi^{(3)}$ respectively, which generate optical polarizations that scale linearly and cubically with excitation field strength:
\begin{align}
    \nonumber P(t) &= \epsilon_0\left[\chi^{(1)}E(t) + \chi^{(3)}E^3(t) + \dots\right]\\
    &= P^{(1)}(t) + P^{(3)}(t) + \dots
\end{align}
where for simplicity we have taken the polarization and fields to be scalar quantities. As shown in Fig.~\ref{Fig1}a, each polarization term gives rise to distinct emitted fields that we call the linear signal $E_{\text{Linear}}$ (which underlies optical absorption and fluorescence) and the four-wave mixing (FWM) signal $E_{\text{FWM}}$ respectively. 

To describe time-domain spectroscopic measurements, it is often convenient to recast the nonlinear polarization in terms of optical response functions $S^{(n)}$:
\begin{align}
    P^{(1)}(t) &= \int^t_{-\infty} S^{(1)}(t_1)E(t-t_1)dt_1\\
    \nonumber P^{(3)}(t) &= \int^t_{-\infty}\int^{t_3}_{-\infty}\int^{t_2}_{-\infty}S^{(3)}(t_3,t_2,t_1)\\
    &\hspace{1.5cm} E(t-t_3)E(t_3-t_2)E(t_2-t_1)dt_1dt_2dt_3.
\end{align}
It can be shown straightforwardly that for impulsive excitation by delta function pulses, each polarization term $P^{(n)}$ becomes identical to its respective optical response function $S^{(n)}$ with the appropriate time-arguments \cite{Mukamel1999,Yang2007,Seibt2013-1}. In this time-domain picture, the (non-)linear optical response function may be interpreted in terms of sequential changes in the system's density matrix induced by interactions with each incident optical field. These sets of changes are collectively termed quantum pathways, and may be represented by so-called (double-sided) Feynman diagrams \cite{Hamm2011}. Feynman diagrams consist of ladders of density matrix elements that begin and end in a population state. Time advances upwards, and arrows pointing (out)inward with respect to each diagram (de-)excite the bra or ket of the appropriate density matrix element. We present examples of such diagrams in Fig.~\ref{Fig1} and discuss their interpretation below, but for a more detailed discussion and rules on their construction we refer readers to more complete reference texts \cite{Mukamel1999,Hamm2011}.

\subsection{Linear Absorption and Fluorescence}

Feynman diagrams representing absorption and fluorescence quantum pathways are shown in Figs.~\ref{Fig1}b and \ref{Fig1}c. First, we consider the absorption diagram in Fig.~\ref{Fig1}b, which begins in an initial ground population state $\Ket{g}\Bra{g}$. An incident electric field, represented by the first arrow, converts this population into a coherence $\Ket{e}\Bra{g}$. The last arrow then converts the intermediate coherence into a final excited population state $\Ket{e}\Bra{e}$. We note that this last arrow does not correspond to an excitation field, but instead originates from taking the expectation value of the macroscopic polarization. The fluorescence diagram in Fig.~\ref{Fig1}c, which begins in an initial excited population state, may be interpreted in the same way. Without the need for an incident field however, the physical origin of the interaction by the first arrow is not as clear. We may informally connect this interaction to perturbation by the vacuum field \cite{Milonni1976}, but a proper interpretation is not possible in the current semi-classical picture.

By referring to the Feynman diagram shown in Fig.~\ref{Fig1}b, the linear optical response function may be written as:
\begin{align}
    S^{(1)}(t) = \frac{i}{\hbar}e^{-i\omega_{eg}t - g(t)} + \text{c.c.}
\end{align}
which reflects the dynamics of the coherence $\Ket{e}\Bra{g}$ generated by the first light-matter interaction. These involve oscillation at the energy gap frequency $\omega_{eg}$ and dephasing characterized by the so-called dephasing lineshape function $g(t)$. From here on we neglect the complex conjugate terms of the linear and higher-order response functions due to redundancy. The frequency-domain absorption lineshape is then simply given by Fourier transform of the optical response:
\begin{align}
    \sigma_a(\omega) = \frac{1}{\pi}\text{Re}\int^\infty_0e^{i(\omega - \omega_{eg})t - g(t)}dt
\end{align}
where we have taken the real (absorptive) part of the lineshape that is measured by absorption spectroscopy. 

The fluorescence spectrum is then similarly:
\begin{align}
    \sigma_f(\omega) = \frac{1}{\pi}\text{Re}\int^\infty_0e^{i(\omega - \omega_{eg} + 2\lambda)t - g^*(t)}dt
\end{align}
where $\lambda$ is the vibrational reorganization energy \cite{Mukamel1999}. Both the absorption and fluorescence lineshapes depend intimately on the underlying electron-phonon coupling via $g(t)$, which we will discuss in further detail after incorporating nuclear motion into our system.

\subsection{Nonlinear Four-Wave Mixing}

In the most general FWM experiment, three excitation pulses \{A,B,C\} (shown in Fig.~\ref{Fig1}d) with variable inter-pulse delay and wavevectors $\{{\bf k}_{\text{A}},{\bf k}_{\text{B}},{\bf k}_{\text{C}}\}$ respectively are directed onto a material. FWM signals generated by all three beams together emit into eight distinct phase-matched directions $\pm{\bf k}_{\text{A}}\pm{\bf k}_{\text{B}}\pm{\bf k}_{\text{C}}$, which provides a way to isolate third-order nonlinear optical signals from stronger signals of lower order. A common choice is to collect the FWM signal emitted in the $-{\bf k}_{\text{A}} + {\bf k}_{\text{B}} + {\bf k}_{\text{C}}$ direction, with the rephasing (photon-echo) pulse-ordering shown in Fig.~\ref{Fig1}d. As the name suggests, the rephasing pulse-ordering generates a photon-echo FWM signal that provides access to homogeneous dephasing dynamics of an inhomogeneous distribution of resonance energies \cite{Cundiff2012}.

The two possible rephasing pulse-ordering Feynman diagrams are shown in Fig.~\ref{Fig1}e which feature more interactions, but may be interpreted in an analogous way as for linear response diagrams. The third-order response function is then written as:
\begin{align}
    S^{(3)}(t,T,\tau) = \left(\frac{i}{\hbar}\right)^3\left[R_{\text{ESE}}(t,T,\tau) + R_{\text{GSB}}(t,T,\tau)\right]
\end{align}
where the excited-state emission and ground-state bleach response functions $R_{\text{ESE}}$ and $R_{\text{GSB}}$ (corresponding to the diagrams shown in Figs.~\ref{Fig1}e and \ref{Fig1}f) are defined by:
\begin{align}
    \nonumber R_{\text{ESE}} &= e^{i\omega_{eg}(\tau-t)}e^{-\Gamma \tau/2}e^{-\Gamma t/2}e^{-\Gamma T}\\
    &\hspace{0.9cm}e^{-g^*(\tau) + g(T) - g^*(t) - g(T + t) - g(\tau + T) + g(\tau + T + t)}\\
    \nonumber R_{\text{GSB}} &= e^{i\omega_{eg}(\tau-t)}e^{-\Gamma \tau/2}e^{-\Gamma t/2}e^{-\Gamma T}\\
    &\hspace{0.9cm}e^{-g^*(\tau) + g^*(T) - g(t) - g^*(T + t) - g^*(\tau + T) + g^*(\tau + T + t)}
\end{align}
where $\Gamma = 1/T_1$ is the population relaxation rate of $\Ket{e}\Bra{e}$. Because the excited state lifetime $T_1$ exceeds a nanosecond in SiV$^-$ centers \cite{Rogers2014-2}, population dynamics are negligible in the (sub-)picosecond regime considered here.

\subsection{Electron-Phonon Coupling and the Dephasing Lineshape Function}

In solid-state materials, the Hamiltonian is often partitioned into an electronic system, a thermal reservoir (phonon modes), and their interaction. If the thermal reservoir is taken to be a set of harmonic oscillators, we may adopt the spin-boson Hamiltonian \cite{Leggett1987,Butkus2012}:
\begin{align}
    \nonumber H &= \Ket{e}\Bra{e}(\hbar\omega_{eg} + \lambda) + \sum\limits_\alpha \hbar\omega_\alpha a^\dagger_\alpha a_\alpha\\
    &\hspace{3cm}+ \Ket{e}\Bra{e}\sum\limits_\alpha\sqrt{s_\alpha}\left(a_\alpha^\dagger + a_\alpha\right)
\end{align}
where $\hbar\omega_{eg}$ is the transition energy, $a_\alpha^{(\dagger)}$ are the creation/annihilation operators for phonon mode $\alpha$, and $s_\alpha$ is the Huang-Rhys factor that quantifies the electron-phonon coupling strength for mode $\alpha$. In order, each term represents the energy of the electronic excitation, the energy of the thermal reservoir, and the electron-phonon coupling respectively.

In electronic systems comprised of discrete states, the primary effect of electron-phonon coupling is to modulate transition energies via elastic interactions \cite{Mukamel1999,Liu2019-2}. In this case, the effect of the environment on a given electronic transition is completely characterized by the spectral density $J(\omega)$. Because energy gap modulation comprises the microscopic origin of coherence dephasing, the dephasing lineshape function may be directly related to the spectral density. It may be shown \cite{Mukamel1999} that for the spin-boson model, the dephasing lineshape function is given exactly by:
\begin{align}
    \nonumber g(t) &= \frac{1}{2\pi}\int^\infty_{-\infty}\left[1 - \cos(\omega t)\right]\coth\left(\frac{\beta\hbar\omega}{2}\right)J(\omega)d\omega\\
    &\hspace{2.1cm} + \frac{i}{2\pi}\int^\infty_{-\infty}\left[\sin(\omega t) - \omega t\right]J(\omega)d\omega
\end{align}
where $\beta = \frac{1}{k_B\mathcal{T}}$ and we have denoted temperature by $\mathcal{T}$ to avoid confusion with the time-delay $T$.

\section{Results and Discussion}

The electronic fine-structure of SiV$^-$ centers in diamond consists of four closely-spaced optical transitions of orthogonal polarizations \cite{Clark1995}. Although SiV$^-$ centers benefit from greatly reduced electron-phonon coupling due to their inversion symmetry, phonon-mediated processes still persist as primary sources of decoherence \cite{Rogers2014-1,Sukachev2017}. Here we consider electronic transitions without any inhomogeneous broadening, which is a good approximation under low-strain conditions \cite{Hepp2014}.

\begin{table}[t]
\caption{\label{Table1} Simulation parameters used for each of the four optical transitions comprising the SiV$^-$ fine-structure. These consist of the transition dipole moment $\mu_{eg}$ and spectral density reorganization energy $\lambda$ and damping rate $\gamma$.}
\begin{ruledtabular}
\begin{tabular}{c | c | c | c | c | c}
$\hbar\omega_{eg} + \lambda$ (meV) & $\mu_{eg}$ & $\lambda_1$ (meV) & $\gamma_1$ & $\lambda_2$ (meV) & $\gamma_2$\\
\hline
1682.09 & 0.25 & 3 & $0.4\omega_1$ & 4 & $0.05\omega_2$\\
1682.28 & 1 & 3 & $0.4\omega_1$ & 4 & $0.03\omega_2$\\
1683.16 & 0.75 & 3 & $0.4\omega_1$ & 4 & $0.03\omega_2$\\
1683.36 & 0.5 & 3 & $0.4\omega_1$ & 4 & $0.05\omega_2$\\
\end{tabular}
\end{ruledtabular}
\end{table}

\subsection{Simulation Parameters}

\begin{figure}[b]
  \centering
  \includegraphics[width=0.5\textwidth]{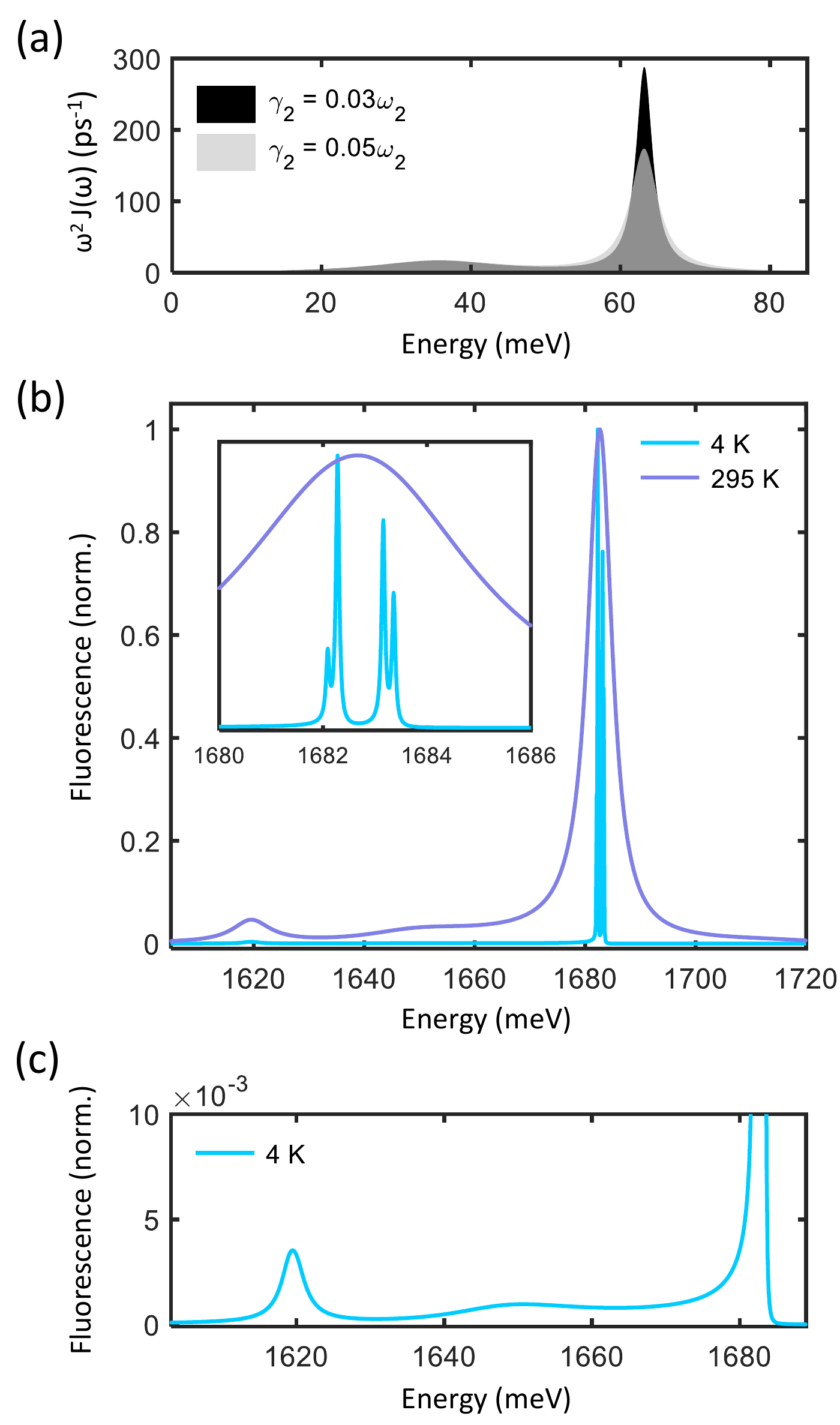}
  \caption{(a) Spectral density calculated according to the parameters given in Table~\ref{Table1}. (b) Fluorescence spectra at 4 K and 295 K simulated for the spectral densities plotted in (a). Inset shows zoom-in of the zero-phonon line, which exhibits fine-structure at 4 K. (c) Magnified plot of the fluorescence spectrum at 4 K in (b) to emphasize the vibronic features.}
  \label{Fig2}
\end{figure}

Two main vibrational modes have been reported in the spectral density: a symmetry-preserving mode of calculated energy $\hbar\omega_1 = 37$ meV and a symmetry-breaking mode of calculated energy $\hbar\omega_2 = 63.19$ meV \cite{Norambuena2016}. The spectral density of a vibrational mode $i$ with energy $\hbar\omega_i$ may be modeled as:
\begin{align}
    J(\omega) = \frac{1}{\omega}\frac{2\sqrt{2}\lambda_i\omega_i^2\gamma_i}{(\omega^2 - \omega_i^2)^2 + 2\gamma_i^2\omega^2}
\end{align}
where $\lambda_i$ is the contribution of mode $i$ to the total reorganization energy $\lambda$ and $\gamma$ is a damping rate that determines the width of vibrational features. We note that the above expression is derived for an exponentially decaying energy gap time-correlation function \cite{Butkus2012}. Here, we model coupling of all four optical transitions, with transition dipole moments $\mu_{eg}$, to these two vibrational modes. Since experimental characterization of the SiV$^-$ spectral density has not yet been performed, simulation parameters are chosen (listed in Table~\ref{Table1}) to agree with experimental fluorescence spectra \cite{Dietrich2014}. To investigate the situation of varying electron-phonon coupling between different fine-structure transitions, we assume marginally different damping rates $\gamma_2 = 0.03\omega_2$ and $\gamma_2 = 0.05\omega_2$ between the orthogonally-polarized doublets of transition energies $\{1682.28,1683.16\}$ meV and $\{1682.09,1683.36\}$ meV respectively. The total spectral densities for each doublet are plotted in Fig.~\ref{Fig2}a, each exhibiting two peaks centered at the vibrational energies $\hbar\omega_1$ and $\hbar\omega_2$. We note that the more strongly damped case $\gamma_2 = 0.05\omega_2$ corresponds to a broader peak in the spectral density.

\begin{figure*}[t]
  \centering
  \includegraphics[width=1\textwidth]{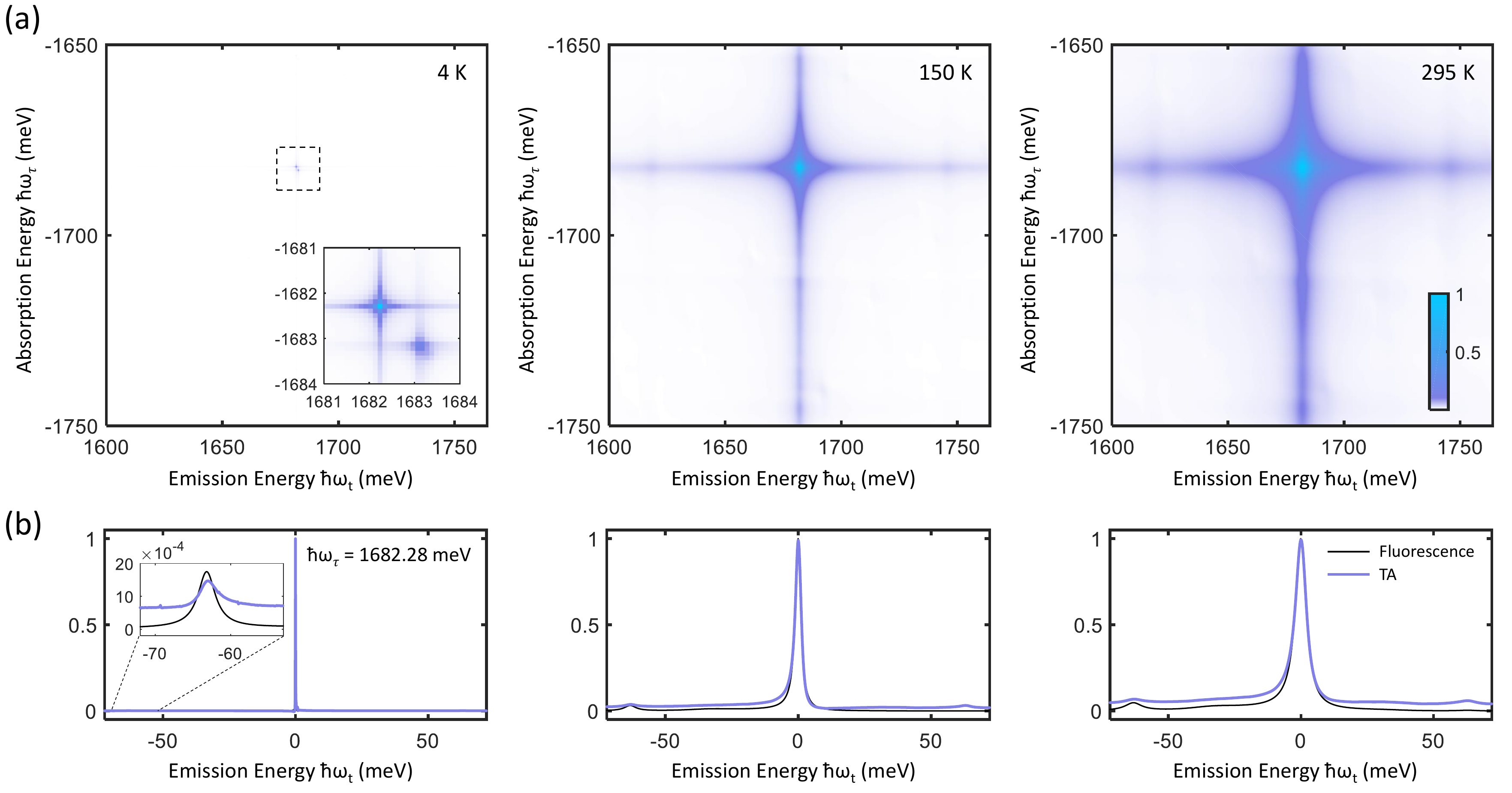}
  \caption{(a) Simulated magnitude rephasing one-quantum spectra spectra for $T = 1$ ps and temperatures 4 K, 150 K, and 295 K as indicated. Inset of 4 K spectrum shows zoom-in of dashed box in main plot, showing multiple peaks arising from the SiV$^-$ fine-structure. We note that coupling between fine-structure transitions is neglected here. (b) Comparison of TA and fluorescence spectra at the same temperatures in (a) plotted as a function of detuning from the transition energy. Inset of the 4 K plot shows zoom-in of the vibronic features located at $-\omega_2$.}
  \label{Fig3}
\end{figure*}

\subsection{Fluorescence Spectroscopy}

Fluorescence spectra simulated with the parameters given in Table~\ref{Table1} are plotted in Fig.~\ref{Fig2}b at cryogenic (4 K) and room (295 K) temperature, which compare well to experiment \cite{Dietrich2014}. Their counterpart absorption spectra are not shown here, but are simply mirror-images of the fluorescence spectra around $\hbar\omega_{eg} + \lambda$ \cite{Mukamel1999}. As shown inset, the SiV$^-$ fine-structure is resolved at 4 K, but becomes indistinguishable at higher temperatures due to thermal broadening of the homogeneous linewidths. Vibronic features are also apparent in the room-temperature fluorescence at lower emission energies, directly corresponding to the two-peak spectral density shown in Fig.~\ref{Fig2}a. To examine vibronic features of the fluorescence spectrum at 4~K, a magnified spectrum is shown in Fig.~\ref{Fig2}c as indicated. Although the same, albeit weaker, vibronic features are observed, individual contributions from each of the fine-structure transitions are indistinguishable. Because the widths of vibronic features are determined not by the electronic dephasing rate, but by the much faster vibronic dephasing rate corresponding to the spectral density width, more sophisticated spectroscopic techniques are necessary to characterize electron-phonon coupling of individual fine-structure transitions in SiV$^-$ center ensembles. In the next sections we show how MDCS, a class of nonlinear spectroscopic techniques, is capable of simultaneously characterizing all fine-structure transitions and their associated vibronic coherence times.

\subsection{Two-Dimensional Coherent Spectroscopy (2DCS)}

As a system possessing inversion symmetry, SiV$^-$ centers do not exhibit a second-order nonlinear optical response. Thus, the lowest-order nonlinear spectroscopies applicable to SiV$^-$ centers are third-order four-wave mixing techniques that measure portions of the third-order optical response function $S^{(3)}$ in the time- or frequency-domains (related by Fourier transform). One such technique is two-dimensional coherent spectroscopy (2DCS), a subset of the broad class of MDCS techniques, which is capable of measuring entire cross-sections of the complex-valued nonlinear optical response via heterodyne detection of nonlinear optical signals. Most commonly, so-called one-quantum spectra $S^{(3)}(\omega_t,T,\omega_\tau)$ are measured that correlate absorption and emission dynamics of a system.

Magnitude one-quantum spectra $\left|S^{(3)}(\omega_t,T,\omega_\tau)\right|$ simulated for a rephasing pulse-ordering are shown in Fig.~\ref{Fig3}a for three temperatures 4 K, 150 K, and 295 K (room-temperature). At 4 K multiple peaks arising from the SiV$^-$ fine-structure are visible, which become indistinguishable at higher temperatures due to thermal broadening. At the higher temperatures 150 K and 295 K, features at $(\omega_t,\omega_\tau) = (\omega_{eg} \pm \omega_2$, -$\omega_{eg})$ and $(\omega_{eg}$, -$\omega_{eg} - \omega_2)$ are visible which indicate coherent electron-phonon coupling. We note that for stronger electron-phonon coupling, additional peaks at $(\omega_t,\omega_\tau) = (\omega_{eg} \pm \omega_2$, -$\omega_{eg} - \omega_2)$ would appear as well.

\begin{figure}[t]
  \centering
  \includegraphics[width=0.5\textwidth]{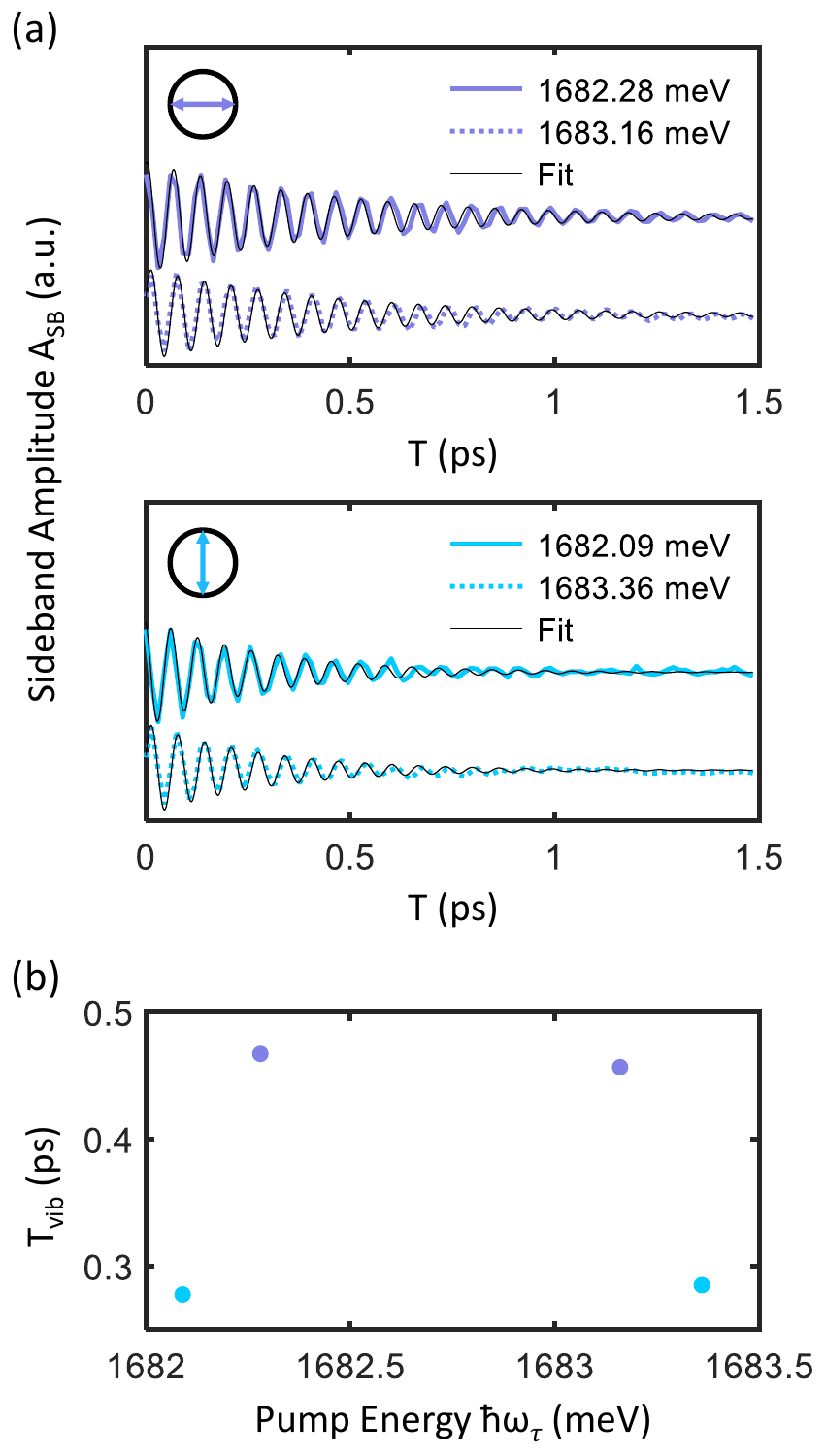}
  \caption{(a) Integrated $\hbar\omega_t = -\omega_2$ sideband intensity evolution at 4 K of horizontally- (top) and vertically-polarized (bottom) transitions as a function of waiting time $T$. (b) Fitted values of the vibronic coherence time $T_{\text{vib}}$ for each transition.}
  \label{Fig4}
\end{figure}

\subsection{Transient-Absorption Spectroscopy (TA)}

A more common third-order nonlinear spectroscopic technique is transient-absorption spectroscopy (TA). Sometimes referred to as transient-SHB \cite{Hochstrasser2007}, TA may be considered a generalization of SHB that resolves temporal population dynamics \cite{Bartels1997}. TA spectra may be acquired via initial excitation by a pump pulse tuned to an electronic resonance, followed by spectrally-resolved measurement of pump-induced changes in absorption of a broadband probe pulse. Experimentally, there is a trade-off between the temporal and frequency resolution of the pump pulse due to its time-bandwidth product. Here we analyze horizontal line-outs of real-quadrature one-quantum spectra Re$\left\{S^{(3)}(\omega_t,T,\omega_{eg})\right\}$ at a given waiting time $T$, which are ideal TA spectra with potentially infinite time and frequency resolution. TA spectra, taken as line-outs from simulated real-quadrature one-quantum spectra, are compared to fluorescence spectra at the corresponding temperatures in Fig.~\ref{Fig3}b. The features of fluorescence and TA spectra are observed to be very similar, consisting of vibronic sidebands at energy $-\hbar\omega_2$, except for an additional sideband at energy +$\hbar\omega_2$ in the TA spectra. However, the true utility of TA spectra becomes apparent once we consider waiting time dynamics during delay $T$.

First, we examine the sideband amplitude at the frequency of the symmetry-breaking vibration, $\hbar\omega_t = -\hbar\omega_2$, sideband amplitude as a function of $T$. At low temperatures the four transitions of the SiV$^-$ fine-structure are clearly resolved, so we plot resultant dynamics from pumping at each of the four transition energies at 4 K in Fig.~\ref{Fig4}a. Coherent oscillations at the phonon frequency $\omega_2$ are observed, which are damped on the sub-ps timescale. We group the plots by emission polarization as indicated, and fit each curve to a damped sinusoid:
\begin{align}
    A_{\text{SB}} = \cos(\omega_2T + \phi)e^{-T/T_{\text{vib}}}
\end{align}
where $T_{\text{vib}}$ is the vibronic coherence time and $\phi$ is the phase of the oscillation. Fitted values of $T_{\text{vib}}$ for each transition are shown in Fig.~\ref{Fig4}b, which exhibit (longer)shorter vibronic coherence times for the (horizontally-)vertically-polarized transitions in accordance with their differing damping rates $\gamma_2 = 0.03\omega_2$ and $0.05\omega_2$ respectively. 

\begin{figure}[b]
  \centering
  \includegraphics[width=0.5\textwidth]{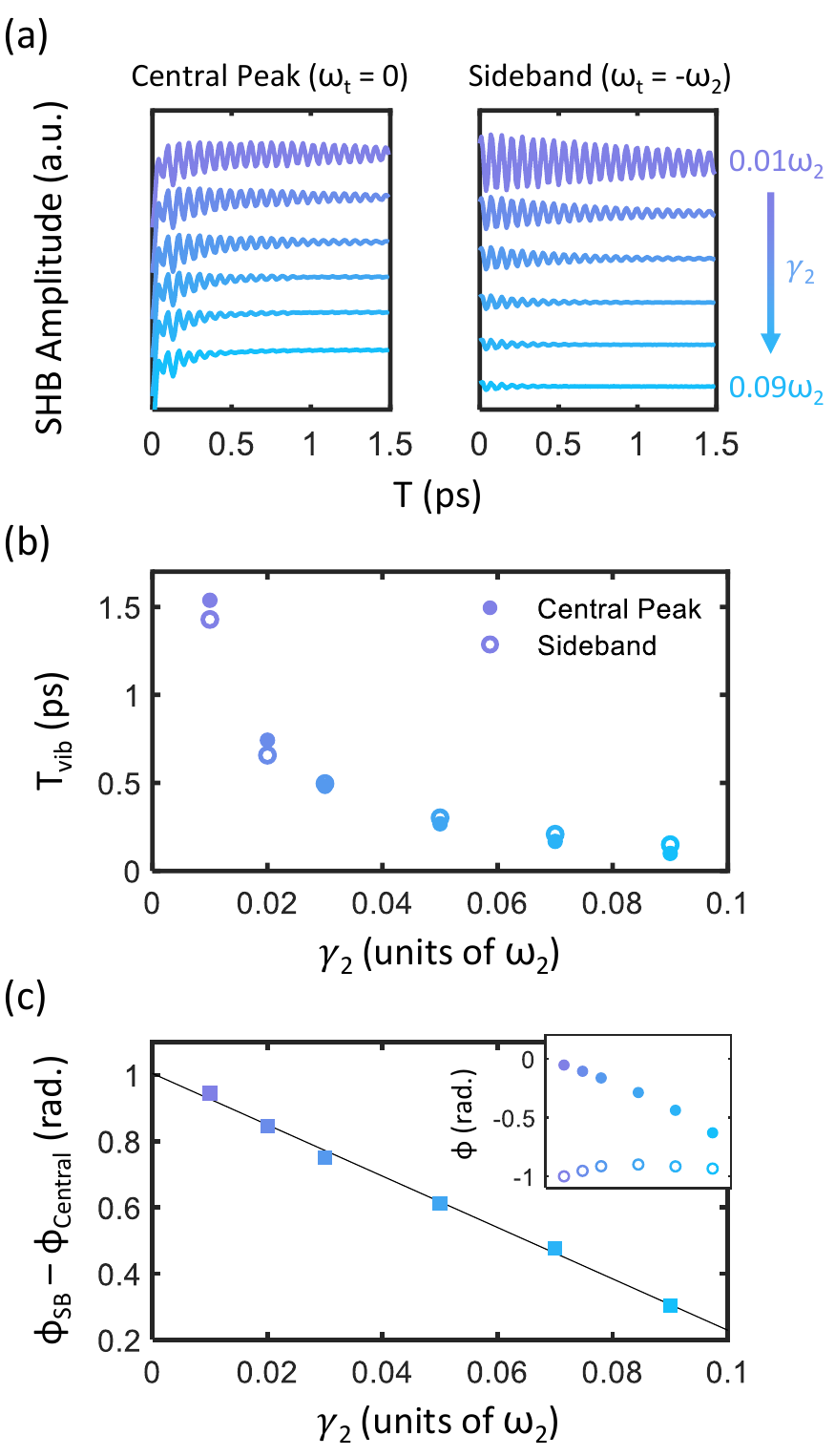}
  \caption{(a) Evolution of the integrated central peak and sideband intensities at 4 K as a function of $T$ for $\gamma_2$ varying from 0.01$\omega_2$ (top curves) to 0.09$\omega_2$ (bottom curves). (b) Values of $T_{\text{vib}}$ fitted from the curves in (a). (c) Phase difference between sideband and central peak oscillations. Inset shows fitted values of the central peak (filled circles) and sideband (open circles) oscillation phases.}
  \label{Fig5}
\end{figure}

We next investigate the general dependence of vibronic oscillations in SHB spectra on the damping parameter $\gamma_2$. The integrated central peak and sideband evolutions are plotted in Fig.~\ref{Fig5}a for $\gamma_2$ increasing from 0.01$\omega_2$ to 0.09$\omega_2$. Fitted values of $T_{\text{vib}}$ for both peaks are then plotted in Fig.~\ref{Fig5}b, which agree well. One may thus relate measured values of $T_{\text{vib}}$ to an underlying damping rate $\gamma_2$ via comparison to simulation. A complementary quantity that may be used to quantify $\gamma_2$ is the oscillation phase. In Fig.~\ref{Fig5}c the difference between the sideband phase $\phi_{\text{SB}}$ and central peak phase $\phi_{\text{Central}}$ is plotted, which decreases linearly with increasing $\gamma_2$. This quantity is more robust than the individually-fitted phases $\phi_{\text{SB}}$ and $\phi_{\text{Central}}$ (plotted inset), which are susceptible to changes in the global phase of the nonlinear signal \cite{Bristow2008}.

\subsection{Three-Dimensional Coherent Spectroscopy (3DCS)}

\begin{figure*}[t]
  \centering
  \includegraphics[width=1\textwidth]{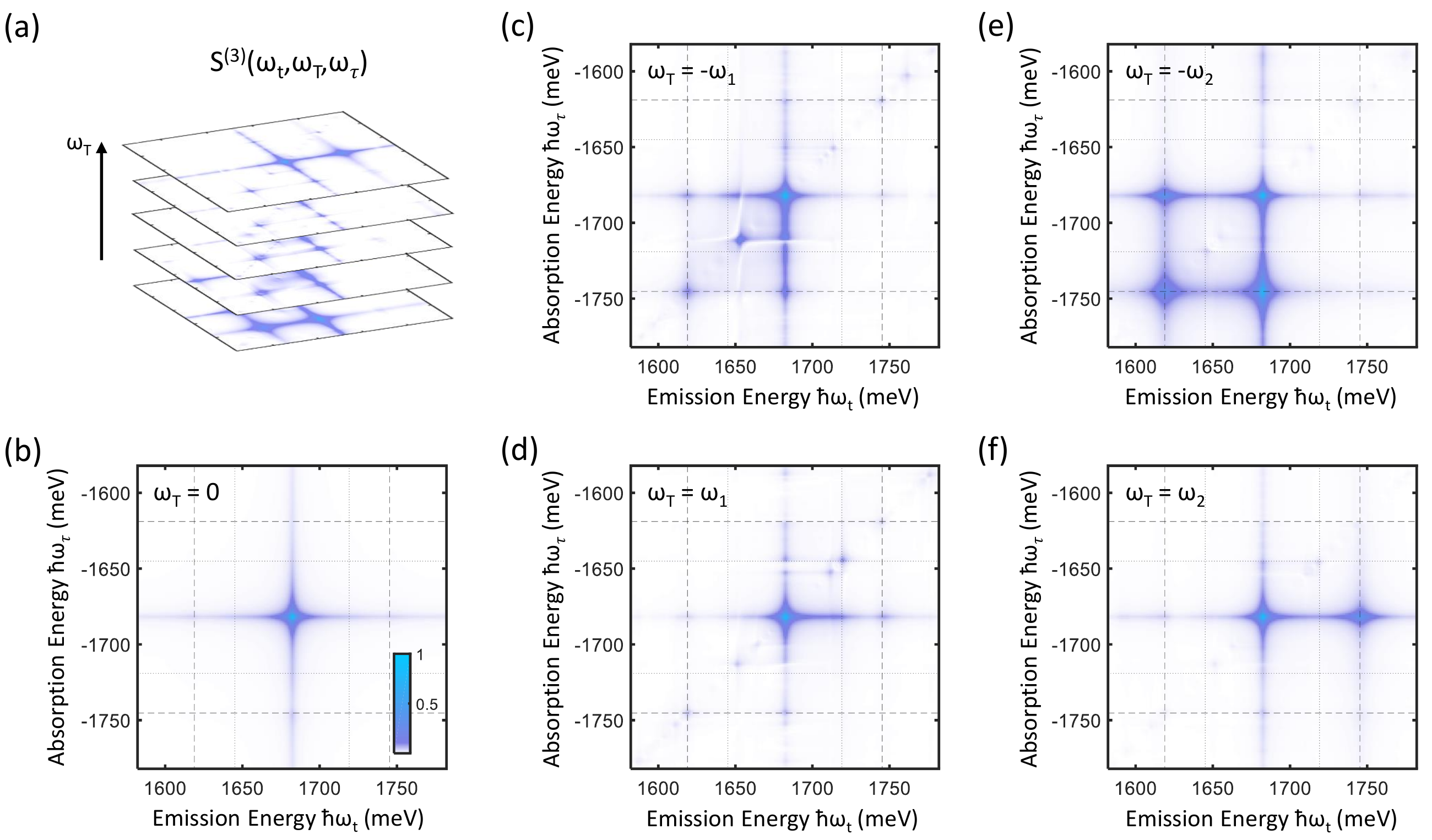}
  \caption{(a) Schematic of three-dimensional spectral solid composed of coherence map cross-sections along $\omega_T$. (b-f) Magnitude coherence maps taken at 150 K along $\omega_T = 0$, $\pm \omega_1$, and $\pm \omega_2$ as indicated. Dotted and dashed lines in (c,d) and (e,f) indicate locations of $\hbar\omega_{eg} \pm \omega_1$ and $\hbar\omega_{eg} \pm \omega_2$ respectively.}
  \label{Fig6}
\end{figure*}

We now discuss the most general MDCS technique, three-dimensional coherent spectroscopy (3DCS), capable of extracting the maximum amount of information from a system's third-order optical response \cite{Li2013,Cundiff2014}. 3DCS involves acquisition of three-dimensional spectral solids along absorption and emission energy axes $\hbar\omega_\tau$ and $\hbar\omega_t$ in addition to a third mixing energy axis $\hbar\omega_T$. As the notation suggests, 3DCS is performed by acquisition of one-quantum spectra as a function of delay $T$ and subsequent Fourier transform. By taking cross-sections of the three-dimensional spectral solid, called coherence maps (shown in Fig.~\ref{Fig6}a), vibronic signatures may be isolated with exceptional clarity \cite{Policht2018}.

In Figs.~\ref{Fig6}b-f, coherence maps taken along $\omega_T = 0$ and the vibrational frequencies $\pm \omega_1$ and $\pm \omega_2$ are plotted. First, the zero-frequency coherence map shown in Fig.~\ref{Fig6}b is primarily composed of a central peak at $(\omega_\tau,\omega_t) = (-\omega_{eg},\omega_{eg})$ with minimal vibronic signatures. Zero-frequency coherence maps are therefore ideal for isolating dephasing and relaxation dynamics of bare electronic transitions. Next, coherence maps along $\omega_T = \pm \omega_1$ are plotted in Fig.~\ref{Fig6}c and \ref{Fig6}d, in which weak vibronic features appear adjacent to the central $(\omega_\tau,\omega_t) = (-\omega_{eg},\omega_{eg})$ peak at absorption and emission frequencies $\omega_{eg} \pm \omega_1$ (indicated by the dotted lines). In contrast, coherence maps along $\omega_T = \pm \omega_2$ exhibit extremely strong vibronic resonances at absorption and emission frequencies $\omega_{eg} \pm \omega_2$ in distinct patterns. 

\begin{figure}[t]
    \centering
    \includegraphics[width=0.5\textwidth]{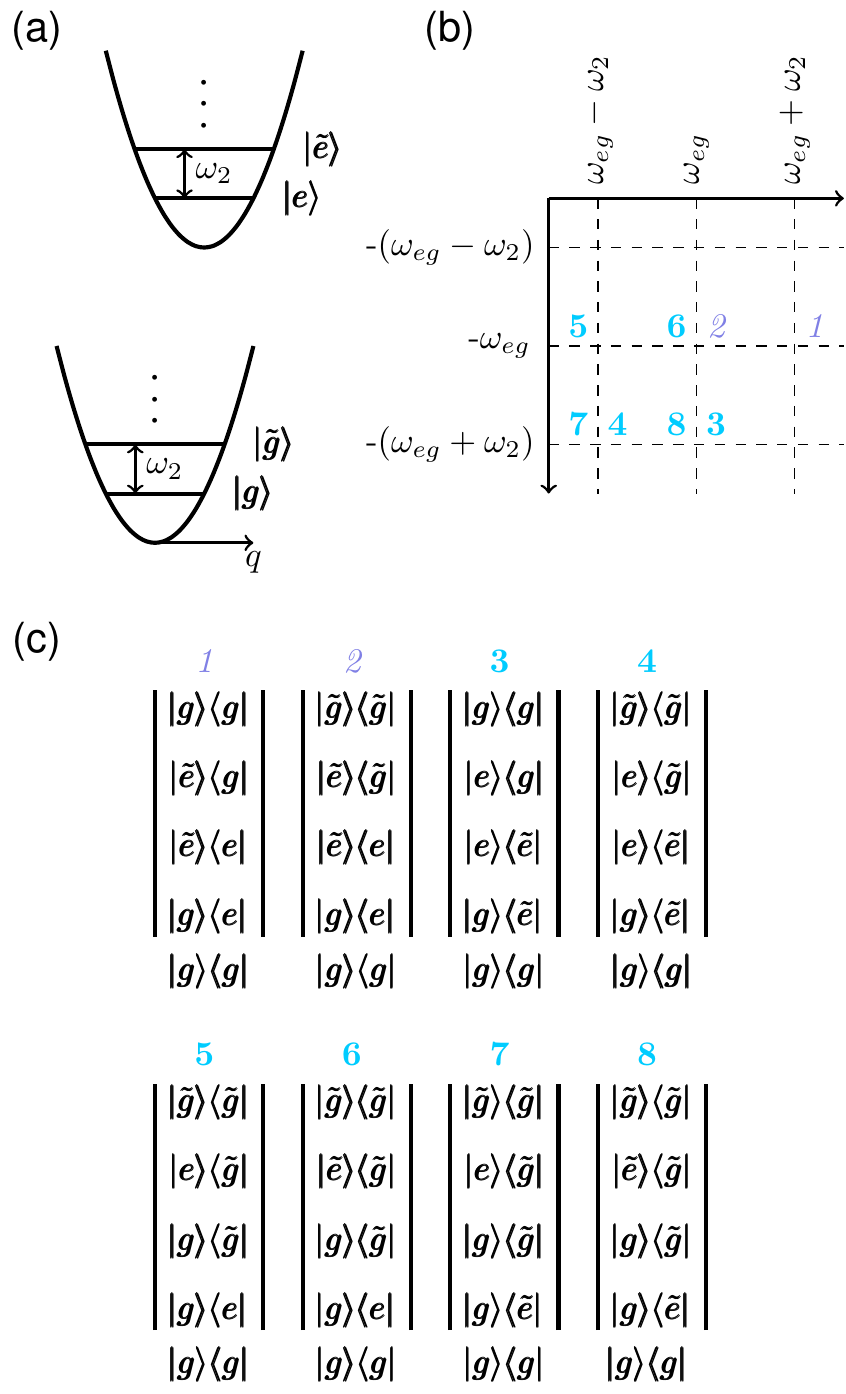}
    \caption{(a) Displaced oscillator model consisting of ground and excited electronic state harmonic potentials along a vibrational coordinate $q$. (b) Schematic coherence map showing peak positions of each numbered Feynman diagram. (c) Feynman diagrams involving intermediate vibrational coherences during delay $T$. Purple (italicized) and blue (non-italicized) numbers indicate positive and negative frequency vibrational coherences respectively.}
    \label{Fig7}
\end{figure}

To understand the coherence map peak structure in Figs.~\ref{Fig6}e and \ref{Fig6}f, we model our system using an equivalent displaced oscillator model in which transitions occur between ground and excited ladders of states separated by the vibrational energy $\hbar\omega_2$ (shown in Fig.~\ref{Fig7}a). Transition dipole moments between initial and final states of $m$ and $n$ vibrational excitations respectively are proportional to the vibrational wavefunction overlap integral \cite{deJong2015}:
\begin{equation}\label{FranckCondon}
    F^m_n = e^{-s_2}s_2^{n-m}\left(\frac{m!}{n!}\right)L^{n-m}_m(s_2)^2
\end{equation}
where $L^{n-m}_m$ are the associated Laguerre polynomials. Because the overlap integral decreases with increasing vibrational excitation number (for $s_2 < 1$, which is the case here), we retain only the lowest two states in each ground and excited state manifold $\{\Ket{g},\Ket{\tilde{g}}$ and $\{\Ket{e},\Ket{\tilde{e}}$ respectively. 

For the effective 4-level system shown in Fig.~\ref{Fig7}a, absorption and emission can occur at the transition frequency $\omega_{eg}$ in addition to $\omega_{eg} \pm \omega_2$ (indicated in Fig.~\ref{Fig7}b). More specifically, if the initial density matrix element is a ground state population $\Ket{g}\Bra{g}$ there are eight total third-order quantum pathways which involve an intermediate vibrational coherence during delay $T$. The eight numbered diagrams are shown in Fig.~\ref{Fig7}c, with their corresponding peak positions indicated in Fig.~\ref{Fig7}b. We may then directly relate the peak structure predicted by this diagrammatic approach to the peaks observed in Figs.~\ref{Fig6}e and \ref{Fig6}f. The negative mixing frequency diagrams (3-8) and positive mixing frequency diagrams (1-2) appear in a characteristic chair shape pattern \cite{Policht2018} which directly mirrors the peak positions of Figs.~\ref{Fig6}e and \ref{Fig6}f respectively. By taking ratios between coherence map peak intensities, the Huang-Rhys factor may be extracted by summing Feynman diagram amplitudes determined by equation (\ref{FranckCondon}).

Interestingly, the coherence maps shown in Figs.~\ref{Fig6}c and \ref{Fig6}d do not exhibit the same clear vibronic peak structure. This is due to a transition between the different damping regimes of each vibrational mode, defined by a quantity $\frac{2\lambda}{\beta\gamma^2}$. The two limiting cases are the slow decay regime $\frac{2\lambda}{\beta\gamma^2} \gg 1$, in which coherent oscillations of the dephasing lineshape occur and give rise to vibronic sidebands \cite{Liu2019-3}, and the fast decay regime $\frac{2\lambda}{\beta\gamma^2} \ll 1$, in which such oscillations are largely damped and homogeneous broadening of the bare electronic transition occurs \cite{Butkus2012}. We note that to characterize vibrational modes in the fast decay regime, temperature-dependent measurements of the homogeneous linewidth may be performed. Indeed, such experiments performed on single SiV$^-$ centers \cite{Neu2013} have yielded homogeneous linewidths that increase with an apparent activation energy \cite{Singh2013} close to $\hbar\omega_1 = 37$ meV. For the simulation temperature of 150 K in Fig.~\ref{Fig6}, we have $\frac{2\lambda_1}{\beta\gamma_1^2} = 0.36$ (fast decay) and $\frac{2\lambda_2}{\beta\gamma_2^2} = 10.37$ (slow decay) which explains the difference between vibronic signatures of each mode.

We note that recent fluorescence studies of SiV$^-$ centers in nanodiamonds have reported distinct sidebands of undetermined origin, which could be attributed to electron-phonon coupling or strain-shifted electronic transitions \cite{Lindner2018}. By resolving quantum pathways in three dimensions, coherence map analysis may distinguish between electronic and vibronic coherences of comparable energy scales \cite{Seibt2013-2}.

\subsection{Inhomogeneous Broadening}

The simulations performed here have assumed no inhomogeneous broadening of electronic transitions, which is a good assumption under low-strain conditions \cite{Hepp2014}. Under such ideal conditions, linear fluorescence spectra suffice to characterize phonon energies and Huang-Rhys factors in the coherent coupling limit. However, strain is often introduced unavoidably during sample fabrication \cite{Lindner2018} or even intentionally to achieve long coherence times above dilution-refrigerator temperatures \cite{Sohn2018}. Even under low-strain conditions however, weakly-fluorescent SiV$^-$ centers have been found to exhibit differing degrees of inhomogeneous broadening \cite{Smallwood2018}. In these situations, nonlinear spectroscopies such as 2DCS and 3DCS are required to characterize possible strain-dependent phonon energies and coupling strengths of the inhomogeneous distribution \cite{Seibt2013-1,Liu2019-2}.

\section{Conclusion}

In this paper, we have presented simulated optical spectra of SiV$^-$ centers coupled to vibrational modes of discrete energy. We demonstrate that nonlinear spectroscopic techniques, namely transient-absorption and multi-dimensional coherent spectroscopy, may be used to completely characterize the spectral density of coherent coupling to a vibrational mode of energy $\hbar\omega_2$ = 63.19 meV. However, coupling to the other dominant vibrational mode of energy $\hbar\omega_1$ = 37 meV is strongly damped and must be characterized via temperature-dependent linewidth measurements. We also demonstrate the ability of TA and MDCS to resolve differences in electron-phonon coupling between adjacent transitions in the SiV$^-$ fine-structure. Our study provides a roadmap for experimental characterization of electron-phonon coupling in SiV$^-$ centers, which may be translated to other inversion-symmetric color center systems such as germanium \cite{Bhaskar2017} or tin \cite{Trusheim2020} vacancy centers in diamond.

We thank Ronald Ulbricht for careful reading and critique of the manuscript.

\end{document}